\documentstyle[aps,eqsecnum,pre,multicol]{revtex}
\input epsf

\begin{document}
\title{\bf Computer simulation of fluid phase transitions}
\author{Nigel B. Wilding}
\address{Department of Mathematical Sciences, The University of Liverpool,\\
Liverpool L69 7BZ, U.K.}

\tighten
\maketitle

\begin{abstract} 

The task of accurately locating fluid phase boundaries by means of
computer simulation is hampered by problems associated with sampling
both coexisting phases in a single simulation run. We explain the
physical background to these problems and describe how they can be
tackled using a synthesis of biased Monte Carlo sampling and histogram
extrapolation methods, married to a standard fluid simulation algorithm.
It is demonstrated that the combined approach provides a powerful method
for tracing fluid phase boundaries.

\end{abstract}
%\twocolumn

%\draft

%\pacs{PACS numbers: 64.60.Fr, 05.70.Jk, 68.35.Rh, 68.15.+e}

%\begin{multicols}{2}

%\vspace*{-0.5cm} %DELETELINE
\begin{multicols}{2} %DELETELINE

\section{Introduction}

One of the fundamental tasks of statistical mechanics is to forge the
link between the {\em microscopic} (atomic-scale) description of a
particular substance and its equilibrium {\em macroscopic}
(thermodynamic) properties. Typically the former is prescribed in terms
of a {\em model}, ie. a specification of the constituent molecules and
their mutual interactions. Given this, the challenge is to derive the
associated macroscopic properties ---quantities such as the
compressibility, heat capacity and, above all, the phase behaviour i.e.
the conditions under which the substance forms a solid, liquid or gas.
Computer simulation allows one to do this.

In many respects, a simulation can be viewed as an experiment performed
not on a real substance, but on a model system stored in a computer's
memory. As in real experiments, one has to prepare a properly
equilibrated sample under the desired thermodynamic conditions; and, just
as in real experiments, one can measure the physical properties of that
substance, perhaps leading to new discoveries as a result! But, a
simulation can have important advantages over a real experiment. Since
one is dealing with a numerical model, substances can be studied that
are too expensive, too complicated or too dangerous to be tackled by
real experiment. Furthermore, since the simulator has access to full and
complete information about the state of the simulated system, there are
fewer restrictions on just which properties can measured. Accordingly,
information and insight can be gleaned from a simulation which would not
readily be obtainable by experimental means \cite{FOOT0}.

However simulations do have their limitations. The chief drawbacks are
constraints on computer speed and memory; most contemporary
computers can deal only with systems comprising a few thousand
particles --many orders of magnitude fewer than the $\sim 10^{23}$
typically found in experimental samples. Such restrictions engender
so-called finite-size effects in the results, ie. spurious artifacts and
systematic discrepancies (compared to the bulk limit) the magnitude of
which need to be quantified. It should also be borne in mind that
irrespective of the sophistication of the simulation techniques
employed, results for macroscopic equilibrium behaviour will reflect
reality only to the extent that the underlying model correctly captures
the true microscopic nature of the substance under study. As the old
computing adage goes: rubbish in, rubbish out.

Simulation strategies for obtaining the equilibrium phase behaviour of
classical fluid systems come in a profusion of different shapes and
forms, but broadly speaking fall into two main categories: Molecular
Dynamics (MD) and Monte Carlo (MC). Both have been previously discussed
extensively in a number of introductory texts and articles, see eg.
refs. \cite{FRENKEL,ALLEN,GOULD,LANDAU}. The MD approach involves computing
the phase space trajectories of a system of mutually interacting
particles by integrating Newton's equations of motions through
time.  Physical properties are measured in terms of configurational or
time averages as the simulation evolves. MD represents an
attractive method in situations where one is interested in obtaining
dynamical information about a system, and can also be employed to obtain
single phase thermodynamical properties. However for studies of phase
transitions, ie. the process by which one phase spontaneously transforms
into another, it is rarely a suitable approach because of the problem of
hysteresis (superheating and supercooling) wherein the temperature and
pressure at which the transition occurs in a simulation may differ
significantly from that of the bulk system. This matter is described in
detail in sec.~\ref{sec:hysteresis}. 

Phase transitions (it now seems quite generally agreed) are best tackled
by Monte Carlo (MC) simulation. Here one employs a stochastic Markov
process to generate a sequence of equilibrium configurations of the
model system of interest; physical properties are measured as
configurational averages over the sequence \cite{FRENKEL,ALLEN,GOULD,LANDAU}. The
actual mechanisms by which the system evolves from one equilibrium
configuration to the next are essentially {\em artificial}, so
information about physical dynamical processes is strictly limited.
Nevertheless this loss is compensated for by potential gains in a
variety of other areas. Specifically, freed from the strictures imposed
by physical dynamics, the simulator is at liberty to construct any
number of elaborate schemes via which the simulation efficiently
explores the space of possible model configurations. By so doing it
becomes possible to bridge time and length scales which \end{multicols}
\twocolumn \newpage \noindent cannot be probed using MD. 

In this article we shall focus on one area in which recently developed
MC simulation techniques allow one to circumvent an old problem, namely
that of accurately obtaining the phase behaviour of model fluids. The
approach we describe is in essence a synthesis of a number of existing
simulation techniques (developed originally in the context of lattice
spin models) which, when married with a standard grand canonical or
constant pressure ensemble fluid simulation, provide a powerful and
efficient route to phase coexistence properties. Before embarking on
this description, however, it is instructive to review some key aspects
of phase transition phenomenology, in particular the origins of the
hysteresis effect which for many years plagued simulation studies of
phase transition and which necessitated the new methodological advances.

\section{Hysteresis, Interfaces and the Free Energy Barrier}

\label{sec:hysteresis}

The phase diagram of a typical simple substance such as Argon, is
depicted in schematic form in figure 1. Depending on the imposed
conditions of temperature and pressure, the substance can exist in 
three phases: solid, liquid or gas. The corresponding regions of the
phase diagram are delineated by phase boundaries at which transitions
occur from one phase to another. Notable features of this phase diagram
are the triple point where three phase boundary lines meet and the
liquid-gas critical point which terminates the liquid-gas phase boundary
and beyond which all distinction between the liquid and the gas
vanishes.

\begin{figure}
\setlength{\epsfxsize}{6.5cm}
\centerline{\mbox{\epsffile{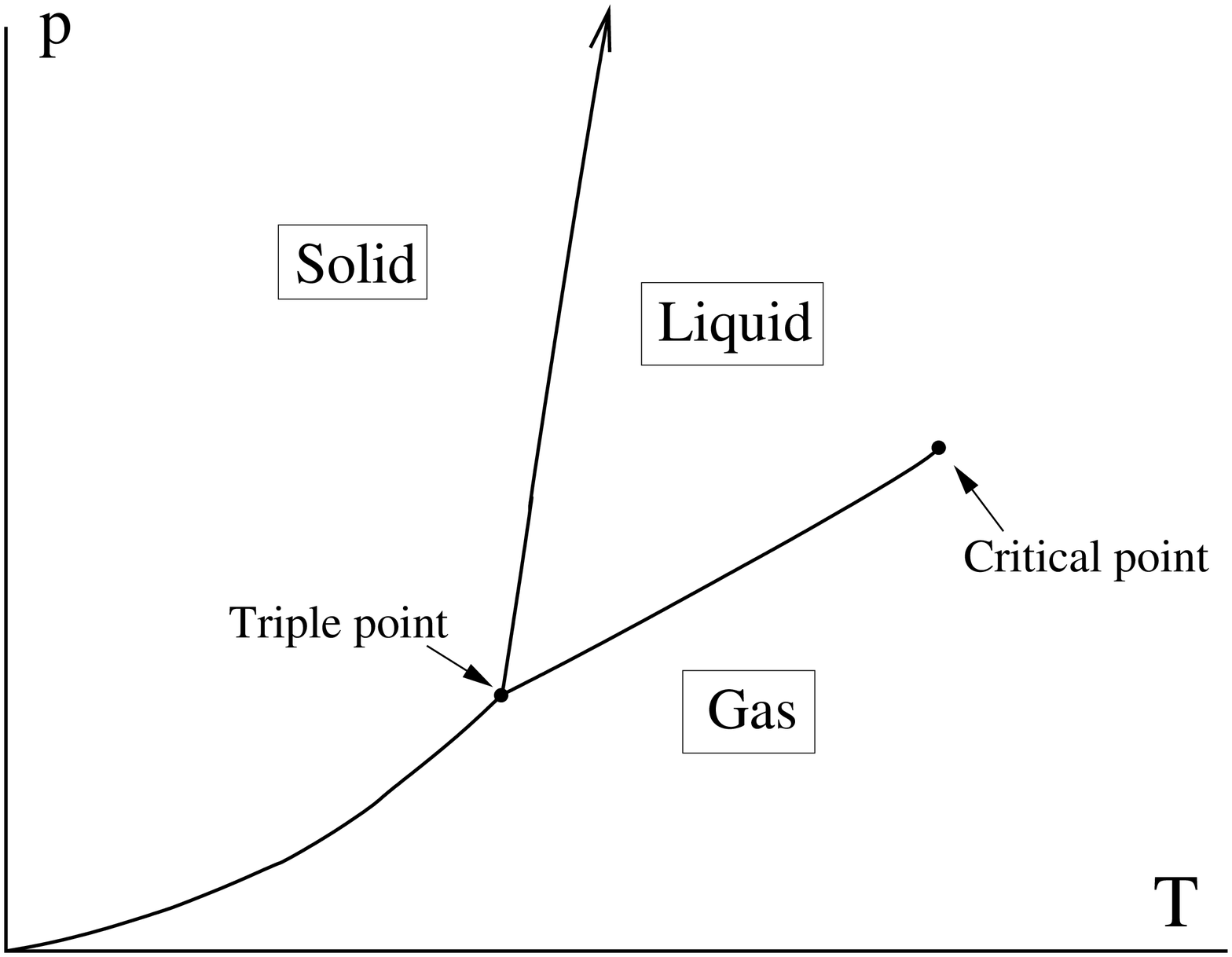}}}
\vspace*{5mm}
\caption{Phase diagram of a simple substance in the
pressure--temperature plane.}
\label{fig:pd}
\end{figure}

Phase diagrams such as that of fig.~\ref{fig:pd} can be interpreted
within the framework of thermodynamics by appeal to the concept of free
energy. A system in equilibrium will always choose its state to be that
for which the free energy is least \cite{PLISCHKE}. Phase boundaries
then arise naturally as that sets of points in the phase diagram for
which two phases have the same free energies, being equally favoured
thermodynamically. In experiments, however, a transition from one stable
phase to another will not always occur exactly on the phase boundary.
Instead one generally encounters `overshoot' or `hysteresis', whereby
the actual transition point depends on the thermodynamic history of the
sample.

\begin{figure}
\setlength{\epsfxsize}{4.5cm}
\centerline{\mbox{\epsffile{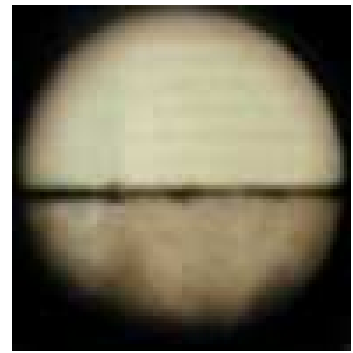}}}
\vspace*{5mm}
\caption{Photograph of coexisting liquid water and steam in a
closed container. The denser liquid occupies the lower portion
of the container, separated by an interface from the vapour.}
\label{fig:coex}
 \end{figure}

The following {\em gedanken} experiment will help to explain the origin
of hysteresis. Imagine we take a purified quantity of a fluid such
as water, place it in a sealed piston-cylinder arrangement at constant
atmospheric pressure, and heat it up very slowly so that it always
remains close to equilibrium. During this process, the water temperature
will rise until at some point it boils --transforming into steam with a
concomitant large and abrupt increase in the system volume. This is an
example of what is called a first order phase transition. If, however,
we stop adding heat before all the water has vapourised we observe {\em
coexistence} between the liquid and vapour phases i.e. a portion of the
container will be occupied by the liquid phase, separated by an
interface from the remainder which contains vapour (cf.
fig.~\ref{fig:coex}). In transforming from one phase to another, a system
always passes through such mixed-phase configurations and it is these
that are responsible for hysteresis. 

It transpires that mixed-phase configurations possess a higher free
energy than pure phase states and since this additional free energy is
wholly associated with the interface itself, it is often referred to as
the {\em surface tension}.  Owing to their surface tension, mixed-phase
configurations are thermodynamically less favourable than pure phase
states at the phase boundary. This has a bearing on phase transitions
such as the vaporisation of our water sample. As the water is slowly
heated it eventually attains the phase boundary temperature $T=373.15$ K
above which steam has a lower free energy than liquid water i.e. becomes
the thermodynamically favoured state. Nevertheless it is possible to
'superheat' liquid water some way beyond this temperature without it
transforming to steam because, in order to do so, it must pass through
the mixed-phase states of higher free energy. A similar effect occurs
when cooling steam --it becomes possible to `supercool' it some way below
$373.15$ K without it liquefying.  Hence the temperature at which the
transition occurs is not that at which the free energies of the two
phases are equal, but instead depends on the initial state of the sample
and the direction and rate in which the boundary is traversed.

Much the same thing occurs in simulations of first order phase
transitions. But here the problem is also intimately bound up with
issues of finite-size effects and simulation timescales. To appreciate
how a simulation behaves near a first order phase transition, it is
instructive to consider a concrete example, namely the liquid-gas
transition of a prototype model fluid --the famous Lennard-Jones fluid
(LJF). Within this model, the interaction potential for two point
particles separated by linear distance $r$ is given by

\[
U(r)=4\epsilon \left( \left(\frac{\sigma}{r}\right)^{12}-
\left(\frac{\sigma}{r}\right)^{6} \right) 
\]
where the parameters $\epsilon$ and $\sigma$ set the strength of the
interaction and the length scale respectively \cite{FRENKEL}.

One way of performing a MC simulation of the LJF is to employ the
grand-canonical ensemble (GCE) in which, for a given system volume $V$, 
the total configurational energy $E$ and particle number $N$ are
permitted to fluctuate stochastically, but with average values determined
by prescribed values of the temperature $T$ and chemical potential
$\mu$. These latter two variables span the phase diagram and by tuning
their values, transitions can be induced between gas, liquid and solid
phases (cf. fig. 1). 

An outline of the operation of a GCE MC simulation is given in box 1. 
Fluctuations in the energy and particle number occur by means of
particle insertions and deletions. MC updates consist of either an
insertion or a deletion attempt, each of which is proposed with equal
probability. For an insertion, one chooses a random position in the
simulation box and calculates the energy change $\Delta E^I$ associated
with placing a new particle at that position. The trial insertion is
accepted with a probability given by a so-called Metropolis rule designed to
ensure that detailed balance is maintained --a necessary condition for
attaining thermodynamic equilibrium \cite{FRENKEL}:

\begin{eqnarray}
p_{\rm acc} (N \to N+1) \hspace*{5cm} \nonumber \\ 
= {\rm min} \left[ 1,\frac{zV}{(N+1)}\exp(-\beta(\Delta E^I))\right ]
\label{eq:met1}
\end{eqnarray}
where $z=\exp(\beta\mu)$ with $\beta=1/k_BT$.

Similarly for a particle deletion, one chooses a particle at random from
those currently present in the system, calculates the energy change
$\Delta E^D$ associated with its removal and then performs the removal
with a probability again given by a Metropolis rule

\begin{eqnarray}
p_{\rm acc} (N \to N-1) \hspace*{5cm} \nonumber \\ 
= {\rm min} \left[ 1,\frac{N}{z V}\exp(-\beta(\Delta E^D))\right ]
\label{eq:met2}
\end{eqnarray}

\setlength{\epsfxsize}{8cm}
\centerline{\mbox{\epsffile{box1.eps}}}
\vspace*{5mm}

{\bf Box 1}: Outline operation of a simple GCE simulation program. For brevity
only the operations for the $x$ coordinate of particle position vector
has been given. Analogous operations apply to the y and z coordinates.
\vspace*{7mm}

Notwithstanding its simplicity, a GCE simulation of this type can
provide a highly efficient computational route to equilibrium fluid 
phase properties. This is especially so if, as is customary, the
interparticle potential $U(r)$ is truncated at some cutoff radius $r_c$
\cite{FRENKEL}. Contributions to $\Delta E^I$ and $\Delta E^D$ then
arise only via local interactions and, by partitioning the simulation
box into cubic cells of side $r_c$, the subset of contributing particles
can be readily identified. Note also, that for studies of liquid-gas
phase transitions one gains nothing by implementing explicit particle
displacement moves in a GCE simulation. Instead these are realized {\em
implicitly} by virtue of repeated particle transfers, sole use of which
not only constitutes a valid algorithm (it is clearly ergodic) but,
moreover, focuses the computational effort on the bottleneck for phase
space evolution namely the  fluctuations in the particle number
density.

\begin{figure}[h]
\setlength{\epsfxsize}{6.5cm}
\centerline{\mbox{\epsffile{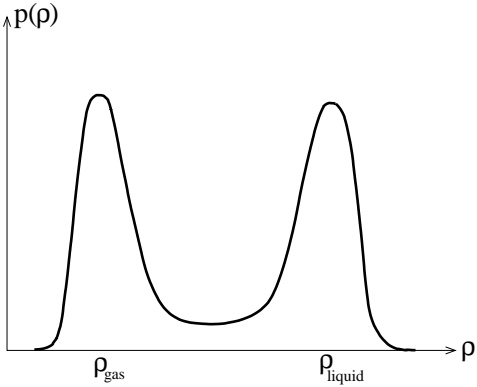}}}
\caption{A schematic diagram of the form of $p(\rho)$ on the liquid-gas phase boundary.}
\label{fig:pbdist}
\end{figure}

Let us now examine how the GCE simulation algorithm of Box 1 behaves in
the vicinity of the liquid-gas phase boundary. The primary observables
of interest from such a simulation are the particle number $N$ and the
configurational energy $E(\{{\bf r}\})$ where $\{{\bf r}\}$ denotes the
set of particle position vectors i.e. the configuration. The
fluctuations of the particle number density, $\rho=N/V$, in particular
provide much insight into the nature of phase coexistence in a
finite-sized system. In a simulation, one can accumulate the probability
density function (pdf) of the number density, $p(\rho)$, in the form of
a histogram averaged over many independent samples of $\rho$. The form
of such a distribution close to liquid-gas coexistence is shown in
schematic form in fig.~\ref{fig:pbdist}.  

The principal feature of this distribution is its bimodal
(double-peaked) character. Each of the two peaks corresponds to one of
the pure phases --the low density peak to the gas phase and the high
density peak to the liquid phase. The location of the liquid-gas phase
boundary line in the $\mu-T$ plane is prescribed formally by the equal
peak weight criteria i.e. by the set of values of $\mu$ and $T$ for
which the integrated weights (areas) under the two peaks are equal. Thus
to locate liquid-gas coexistence in a simulation one must tune $\mu$ and
$T$ until the measured form of $p(\rho)$ is doubly peaked with equal
area under each peak. The problem for simulations is that to obtain
accurate estimates for the relative peak weights, the simulation
procedure must supply bountiful independent samples from each of the two
phases, which in turn necessitates that it pass many times back and
forth between them. 

Unfortunately, the inter-phase route necessarily traverses the
mixed-phase configurations in which regions of both phases coexist
within the simulation box. Such configurations have, on account of their
surface tension, an {\em a-priori} probability that is intrinsically low
compared to those of the pure phase states. This, of course, is the
physical origin of the trough separating the two peaks of $p(\rho)$
which may accordingly be regarded as a ``probability (or free energy)
barrier'' to inter-phase transitions \cite{FOOT1}. The height of this
barrier is taken to be the ratio of the maximum (peak) value of
$p(\rho)$ to its minimum value in the trough. For large barrier heights,
the free energy penalty associated with forming mixed-phase
configurations is so high that transitions between the two pure phases
can become very rare on the timescale accessible to simulation. As a
consequence, the correlation time becomes large, hindering the
accumulation of independent statistics on the relative peak weights and
thence the accurate location of the phase boundary. 

\begin{figure}
\setlength{\epsfxsize}{6.5cm}
\centerline{\mbox{\epsffile{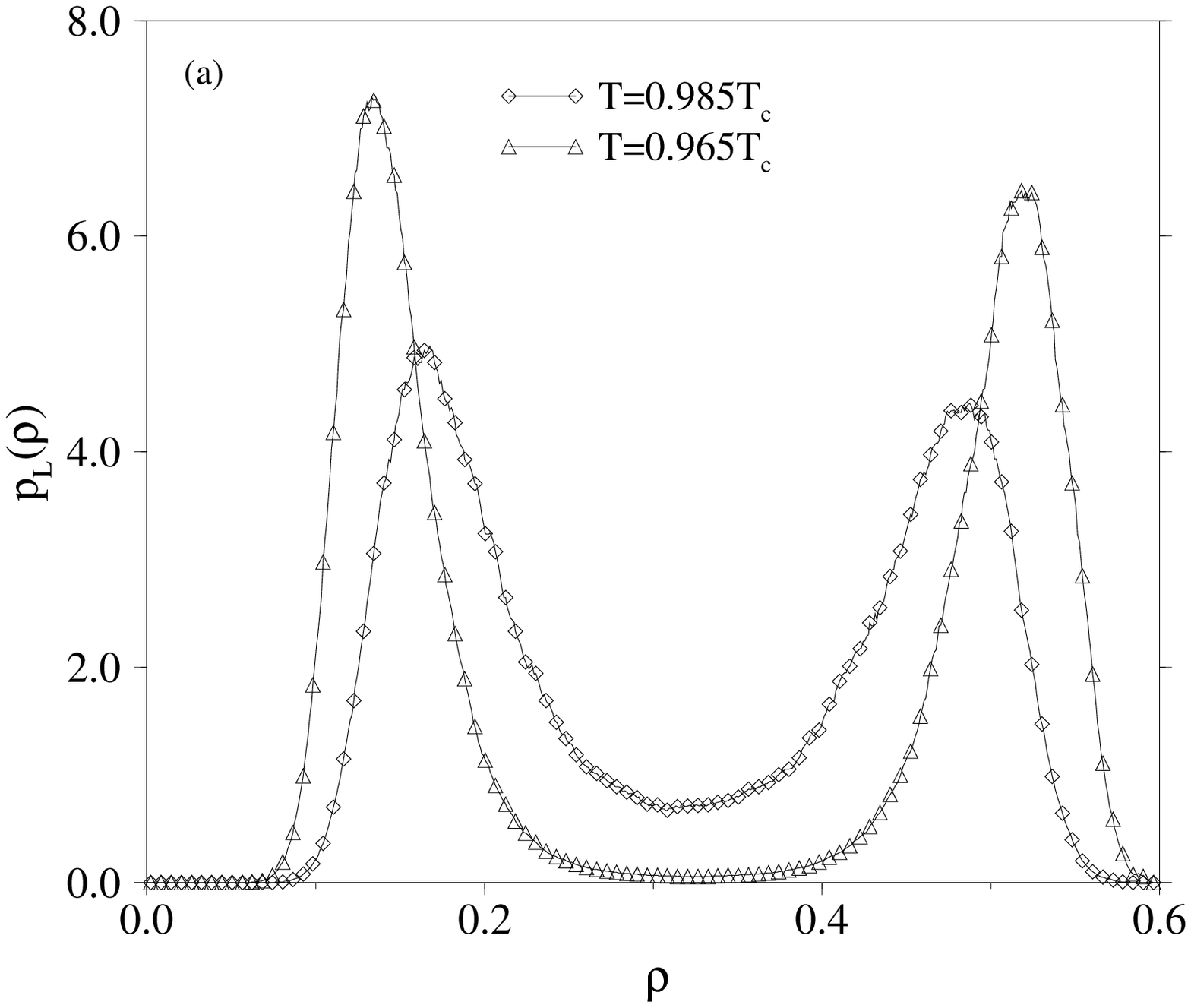}}}
\setlength{\epsfxsize}{7.5cm}
\centerline{\mbox{\epsffile{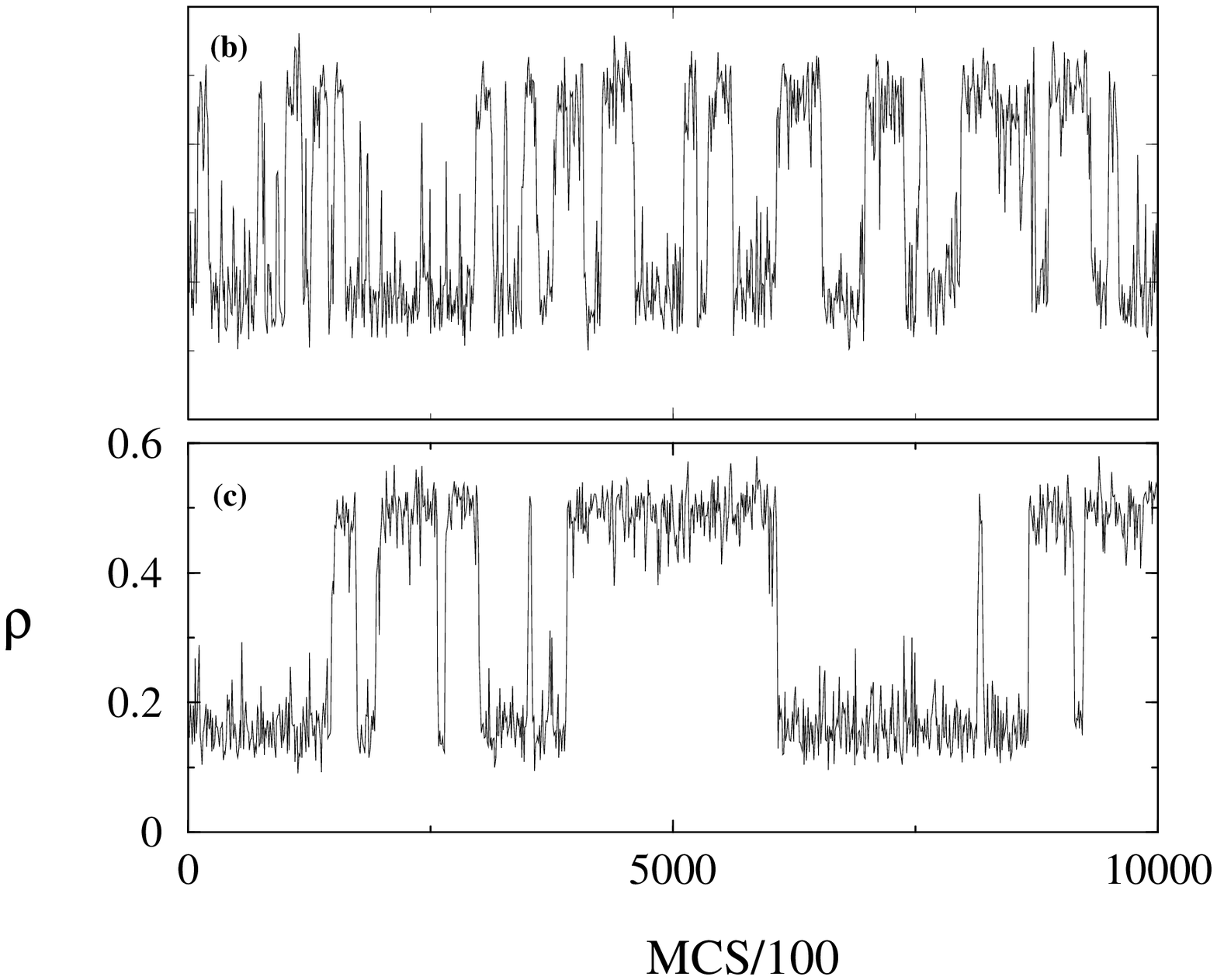}}}
\caption{{\bf (a)} The form of $p(\rho)$ at $T=0.985T_c$ and $T=0.965T_c$
{\bf (b)} The associated time evolution of the number density at
$=0.985T_c$ and {\bf (c)} $T=0.965T_c$}
\label{fig:Tdep}
\end{figure}

The barrier height (and hence the scale of the difficult) depends on the
temperature. At the critical temperature the surface tension vanishes
and with it the probability barrier \cite{FOOT2}. If, however, one
follows the liquid-gas boundary to progressively lower sub-critical
temperatures, thermal fluctuations diminish and the surface tension
grows. This is manifest in a growth in the barrier height accompanied by
a narrowing of the peaks of $p(\rho)$. These features are illustrated in
fig.~\ref{fig:Tdep} which shows the measured form of $p(\rho)$ (obtained
using the GCE algorithm of box 1.) for the LJF close to the phase
boundary for the two temperatures $T=0.985T_c$ and $T=0.965T_c$. Also
shown in fig.~\ref{fig:Tdep}(b) is the corresponding evolution of the
density expressed as a function of the number of Monte Carlo
insertion/deletion attempts. For the higher  temperature, the barrier
height is approximately $5$ and inter-phase transitions are fairly
frequent. On decreasing the temperature to $0.965T_c$, however, the
barrier height increases to a factor $30$ and inter-phase crossings
become comparatively much rarer. This reluctance to transform between
the phases is, of course, none other that the hystereis phenomenon,
viewed not from the standpoint of thermodnamics,  but in terms of the
underlying statistical mechanics \cite{FOOT1}.

Qualitatively similar effects occur if the linear dimension of the
simulation box $L$ is increased. The barrier height grows because the
surface tension of the interface increases proportionally to its area
($\sim L^2$ for $d=3$). Additionally, the peaks of $p(\rho)$ narrow due
to `self-averaging' of fluctuations.  Even for quite modest system sizes
the barrier can be prohibitively large for inter-phase
crossings to occur on the accessible simulation timescale.

Effectively then, one is caught between the `rock' of wishing to
minimise finite-size effects by employing a large simulation box, and
the `hard place' of seeking to sample on timescales  exceeding the
correlation time. Evidently therefore, a superior approach to bare GCE
simulation is called for if one is to tackle the problem of
first order phase transitions at temperatures significantly below that
of criticality. Indeed, over recent years considerable effort has been
invested in developing new MC simulation methods for circumventing the
problems identified above. In the next sections we describe one solution
to the problem which is rapidly becoming the method of choice for high
resolution studies of fluid phase equilibria.

\section{Beating the barrier}

\label{sec:beating}

Approaches for dealing with the free energy barrier in simulations of
phase transitions fall broadly into two categories

\begin{itemize}

\item[(i)] Simulations without interfaces 

\item[(ii)] Biased sampling techniques

\end{itemize}

Foremost in category (i) are methods such as the Gibbs Ensemble Monte
Carlo (GEMC) \cite{PANAGIO87} and Gibbs-Duhem integration \cite{KOFKE},
both of which have enjoyed widespread use and popularity in studies of
phase transitions in model fluids. Although quite distinct in character,
both these methods serve to link the pure phases thermodynamically,
without traversing mixed-phase configurations. Both are versatile and
fairly easy to use. However, they also have significant drawbacks as
discussed in Sec.~\ref{sec:concs}. Specifically, it seems difficult
using the GEMC method to obtain coexistence data of high statistical
quality without a large investment of computational effort. By
comparison, Gibbs-Duhem integration is more efficient, but potentially
suffers from integration errors rendering it difficult to assess the
accuracy of results for phase boundary properties.

More recently, an alternative approach has emerged \cite{WILDING95} --one
which although arguably less straightforward to implement, offers the
rewards of considerably greater efficiency, precision and flexibility
than GEMC or Gibbs-Duhem integration. This scheme is based on a
synthesis of two existing simulation techniques: Multicanonical biased
sampling and Histogram Reweighting. In the remainder of this article, we
shall describe how this combined method operates and set out in a
step-by-step fashion how one goes about implementing it in practice.

\subsection{Multicanonical Sampling}

\label{sec:multi}

Multicanonical MC owes its origin to the biased sampling techniques
first introduced in the 1970's by Torrie and Valleau to calculate free
energies \cite{TORRIE}. Latterly however, such techniques have gained
fresh impetus with the realisation \cite{BERG} that they permit the
bridging of the free energy barrier at a first order phase boundary. In
this context the term ``Multicanonical sampling'' was coined and the
method applied successfully to the study of phase transitions and free
energy landscapes in a variety of lattice-based spin systems
\cite{GUBERNATIS}.

The basic idea underpinning Multicanonical MC is to {\em
preweight} the sampling of configuration space in such a way as to
artificially enhance the occurrence of the mixed-phase configurations of
intrinsically low probability. By so doing it is possible to overcome
the probability barrier separating the two pure phases, thereby allowing
the simulation to pass unhindered between them. The result is a great
reduction in the correlation time of the sampling process. 

Within the GCE framework, the biasing is achieved by use of a
preweighting function incorporated into the Metropolis acceptance
criteria for particle insertions and deletions (cf. Box 1). Its purpose
is to modify (with respect to standard Boltzmann statistics) the
probability with which configurations of the various densities are
visited, in such a way that the measured number density pdf is
approximately {\em flat} over the entire density range separating the
two pure phases. Of course the results of such biased simulations
deviate from Boltzmann statistics and consequently lack direct physical
significance. Nevertheless, it is possible (as we describe below) to
{\em unfold} from the simulation results the unwanted effects of the
imposed bias, thence recovering the physically relevant quantities which
would have been obtained in an unbiased simulation had sufficient
run-time been available. In general, however, there is a price to be
paid for this gain, and that is the expenditure of effort involved in
finding a suitable form for the preweighting function. Fortunately it
transpires that for the purposes of tracing phase boundaries this is not
always necessary, it being possible to obtain a suitable preweighting
function for `free'.

\subsubsection{Formalism}

Let us begin by considering the GCE form of the particle number pdf,
$p(N)$, at inverse temperature $\beta=1/k_BT$ and chemical potential
$\mu$. For a system of volume $V$ this takes the form of  a simple
average of the Boltzmann factor over all possible particle positions
\cite{FRENKEL}:

\begin{equation} 
p(N\;|\;V,\beta,\mu)=\frac{1}{Z}\prod_{i=1}^{N} \left \{ \int_V d r_i \right \} e^{-\beta\cal{H}} \;\;,
\label{eq:pN} 
\end{equation}
where $\cal {H}$ is the configurational Hamiltonian given by ${\cal H}(\{
{\bf r} \},N)\equiv E(\{ {\bf r} \})-\mu N$, while $Z=Z(\beta,\mu)$ is
the partition function which serves to normalise the distribution to unit integrated weight.

The multicanonical method operates by sampling not from a simple
Boltzmann distribution with Hamiltonian ${\cal H}(\{{\bf r}\},N)$, but
from a modified distribution with effective Hamiltonian 

\begin{equation}
\tilde{{\cal H}}={\cal H}+\eta(N)\;,
\label{eq:effham}
\end{equation}
where $\eta(N)$ is a preweighting function defined on the set of particle
numbers $N$ \cite{FOOT3}. The associated particle number distribution is given by
\begin{equation} 
\tilde{p}(N\;|\;V,\beta,\mu,\eta(N))=\frac{1}{\tilde{Z}}\prod_{i=1}^{N} \left \{ \int_V d
r_i \right \} e^{-\beta\tilde{{\cal H}}} 
\label{eq:pNm} 
\end{equation}

Let us now suppose for the sake of argument that we are able to choose
the preweighting function such that $\eta(N)=\ln p(N)$ where $p(N)$ is
the desired Boltzmann density distribution. Inspection of
eqs.~\ref{eq:pN}--\ref{eq:pNm} reveals that this implies $\tilde{p} (N)
= {\rm constant} \; \forall\; N$. To the extent that such a choice of
weight function can actually be realised, the density then performs a
$1$-d random walk over its entire domain, thereby allowing extremely
efficient accumulation of the preweighted histogram $\tilde{p}(N)$.

Unfortunately, this happy state of affairs cannot in general be
immediately achieved because the preweighting function $\eta(N)=\ln
p(N)$ that serves to flatten $\tilde{p}(N)$ is, of course, just the
logarithm of the function we are trying to find! Means must therefore be
found to obtain a form for $\eta(N)$ that approximates $\ln p(N)$
sufficiently well that inter-phase transitions occur with an acceptably
high frequency. More refined forms for $\eta(N)$ can thereafter be
obtained in the further course of the study. 

Let us assume that a suitable preweighting function {\em has} been
found, and a simulation performed to obtain good statistics for
$\tilde{p}(N)$. The next step is to infer the distribution
 $p(N)$ we actually seek by unfolding the effects of the multicanonical
preweighting. This is achievable because knowledge of the preweighting
function tells us exactly by what degree the relative probabilities of
the states of various $N$ were altered in the simulation with respect to
the true Boltzmann statistics. One therefore need only divide out the
relative probability enhancements from $\tilde{p}(N)$ to yield $p(N)$.
This is done for each value of $N$ in the range of interest by the
simple reweighting:

\begin{equation} 
p(N \;|\; V,\beta,\mu)=e^{\eta(N)}\;\tilde{p}( N\; |\;\beta,\mu,\eta(N) ) \;.
\end{equation}

The details of the practical implementation of this procedure are
described in Sec.~\ref{sec:practice}.

\subsection{Histogram reweighting}

\label{sec:histo}

Histogram Reweighting is the second ingredient in our simulation
procedure. It rests upon the observation that histograms of observables
accumulated at one set of model parameters (in our case $\beta$ and
$\mu$) can be analysed to provide estimates of histograms appropriate to
other values of these parameters. Consider the {\em joint} probability
distribution of energy and particle number fluctuations at some
particular parameter values $\beta=\beta_0$ and  $\mu=\mu_0$. Formally
this is given by 

\begin{eqnarray}
p(N,E \;|\; V,\beta_0,\mu_0)\hspace*{5cm}\nonumber \\ 
=\frac{1}{Z_0} \prod_i^N \left \{ \int_V dr_i \right \} \delta(E-E(\{r_i\}))e^{-\beta_0{\cal H}_0}\;,
\end{eqnarray}
where ${\cal H}_0(\{ {\bf r} \},N)\equiv E(\{ {\bf r} \})+\mu_0 N$.
It is easy to show \cite{FERRENBERG} that an estimate for the
form of $p(N,E)$ at some other parameters $\beta=\beta_1,\mu=\mu_1$ can be
obtained from the measured pdf by the simple reweighting:

\begin{equation}
p(N,E\; |\; V,\beta_1,\mu_1)=
\frac{Z_1}{Z_0} e^{-(\beta_1{\cal H}_1-\beta_0{\cal H}_0)} p(N,E\; |\;V,\beta_0,\mu_0)\;,
\label{eq:hr}
\end{equation}
where the ratio $Z_1/Z_0$ is an unimportant constant which is
effectively absorbed into the normalisation. If desired, this
joint distribution can then be marginalised to yield a uni-variable
distribution, eg. the particle number pdf at $\beta_1,\mu_1$:

\begin{equation}
p(N\; | \; V,\beta_1,\mu_1)=\int dE\; p(N,E\;|\; V,\beta_1,\mu_1)\;.
\end{equation}

Hence, in principle at least, a single simulation at one state point in
the phase diagram suffices to obtain information concerning all other
state points. Unfortunately the reality of the situation is less
auspicious. Owing to finite sampling time, it is not possible {\em in
practice} to reweight a single histogram obtained from a simulation at
some $\beta_0,\mu_0$ to arbitrary values of $\beta_1,\mu_1$. Instead the
parameters to which one extrapolates must be close (in a sense we shall
describe) to those at which the simulation was actually performed,
otherwise the procedure loses accuracy. 

The problem is traceable to the fact that the reweighting represented by
eq.~\ref{eq:hr} may drastically modify the relative statistical weights
of the various members of the set of configurations that contribute to
the spectrum of measurements.  Specifically, difficulties arise with
that subset of sampled configurations having very low Boltzmann weight
at the simulation parameters $\beta_0,\mu_0$, members of which
consequently occur only very rarely in the sample. For expectation
values of observables calculated at $\beta_0,\mu_0$, contributions from
this rare subset of configurations do not contribute disproportionately
to statistical uncertainties. However, under the reweighting to
$\beta_1,\mu_1$, the configurations in question may be assigned a much
greater statistical weight, one which does not reflect their actual
representation in the overall sample. This has the effect of magnifying
the overall statistical error on measured expectation values and is
manifest as a reweighted histogram that appears `ragged' in its extremal
regions.

One way of dealing with this problem is to perform a sequence of
separate simulations at strategic intervals across the range of model
parameters of interest. Typically the intervals are chosen such that
histogram of some observable (eg. the energy) accumulated at
neighbouring state points in the sequence overlap within some region of
their domain. The role of Histogram Reweighting to then to interpolate
into the regions of parameter space between the simulation points. In
this context it should be noted that it is possible to combine (in a
self consistent fashion) the results of a number of different
simulations at different model parameters and perform Histogram
Reweighting on the {\em aggregate} data. For a description of this more
sophisticated procedure we refer the reader to
refs.~\cite{FERRENBERG,FRENKEL}.

\subsection{Stitching together the pieces}

\label{sec:practice}

So how does one implement the above formalism in a GCE simulation
of a fluid? In fact the task falls naturally into two parts: performing
the actual multicanonically preweighted simulation and the subsequent
data analysis. We consider them in turn.

The business of implementing the multicanonical preweighting within a
GCE simulation is basically quite straightforward. Assuming an
appropriate set of multicanonical weights has been found, all one need
do is read them in and store them in an array.  The simulation then
proceeds as outlined in Box 1., except that the Metropolis acceptance
probabilities for insertion and deletion (eqns. ~\ref{eq:met1} and
~\ref{eq:met2}) are modified to read: 

\begin{eqnarray}
p_{\rm acc} (N \to N+1) \hspace*{5cm} \nonumber \\ 
= {\rm min} \left[ 1,\frac{\eta(N)}{\eta(N+1)}\frac{zV}{(N+1)}\exp(-\beta(\Delta E^I))\right ]\;,;
\end{eqnarray}

\begin{eqnarray}
p_{\rm acc} (N \to N-1) \hspace*{5cm} \nonumber \\ 
= {\rm min} \left[ 1,\frac{ \eta(N)}{ \eta(N-1)}\frac{N}{z V}\exp(-\beta(\Delta E^D))\right ]\;.
\end{eqnarray}

Consider now the simulation quantities we seek to obtain, namely the
probability distributions of $N, E$ and any other observables of
interest.  Obviously it is tempting to accumulate these distributions in
the form of histograms built up in the course of the simulation by
simply bining successive measurements into an array. But for continuous
variables such as the energy, this strategy necessitates a prior choice
for the histogram bin-width. Should a choice be made that subsequently
turns out to be unsatisfactory, then one is faced with little
alternative but to repeat the whole simulation. A superior approach,
retaining complete information, involves decoupling the data analysis
from the simulation by recording the full history of raw data
measurements. This data is then postprocessed by a separate analysis
program. Although such an approach can make for large simulation output
files, it has the overriding advantage of ensuring maximum flexibility
in terms of data analysis.

To facilitate the post processing approach, the raw data should be
accumulated in the form of a list. Suppose we perform a GCE MC
simulation of the LJF at a given $\beta=\beta_0$ and $\mu=\mu_0$,
employing some chosen weight function $\eta(N)$. As the simulation
proceeds we make a succession of measurements (performed at regular
spaced intervals of time) of the observables $E$ and $N$, together with
any other quantities of interest.  Successive measurements of these
observables are gathered into a list by appending them to a file, viz:
 
\begin{center}  \vbox{  $\{ E_0, N_0,O_0,.. \}$ \\
  $\{ E_1, N_1,O_1,..
\}$ \\
  $\{ E_2, N_2,O_2,.. \}$ \\
  . \\
  . \\
 $\{ E_j, N_j,O_j,.. \} $
\\
  . \\
 . \\
  $\{ E_{M}, N_{M},O_{M},.. \} $ \\
  }  \end{center} 
where $j$ indexes the series of $M+1$ measurements and $O$ denotes an
observable of interest. The data list is analysed following the actual
simulation by a separate post-processing program, the task of which is
three fold. Firstly it should remove from the data the unwanted effects of the
preweighting in order to recover the desired Boltzmann distributed
statistical properties. Secondly, it should output pdf's of the
observables of interest in the form of histograms. Thirdly, it should
(if desired) reweight the data to obtain estimates of histograms
appropriate to parameter values different from those at which the
simulation was performed.

It turns out that the operations of histogram reweighting and bias
removal can be accomplished simultaneously because their mathematical
structures are very similar. To do this, one simply runs through the
data list assigning each entry $j$ a statistical weight

\begin{equation}
w_j=e^{[-(\beta_1-\beta_0)E_j+(\beta_1\mu_1-\beta_0\mu_0)N_j+\eta(N_j)]}\;,
\end{equation}
where $\beta_1,\mu_1$ are the parameters to which one wishes to
extrapolate. The complete set of $M+1$ weights
$w_0,w_1,,,w_M$ is then used to construct the reweighted histogram for
some measured observable of interest $O$: 

\begin{equation}
H(O\;|V,\beta_1,\mu_1)=\sum_{j=0}^{M} w_j\delta(O-O_j).
\end{equation}
This histogram, once suitably normalised, constitutes a discrete
estimate for the probability distribution
$p(O^\prime\;|\;V,\beta_1,\mu_1)$. Implicit in its construction is the
specification of the bin-width, which may need to be tuned to strike a
balance between resolution and data smoothness. But with the raw list
safely stored in a file, this is something that can be done very quickly.

\section{Tracing coexistence curves}

We now turn to the actual simulation procedure by which a fluid phase
boundary can be determined. For illustrative purposes, we shall remain with
our prototype example, the liquid-gas transition of the LJF. The essential
approach is nevertheless rather general and can be applied to other
types of phase transition such as demixing transitions in fluid
mixtures.

Clearly one of the key components of our procedure is multicanonical
preweighting, use of which generally entails a degree of preliminary
effort in order to determine a suitable preweighting function. It
transpires, however, that provided one begins tracing the phase boundary
from near the critical point which marks its terminus, no additional
effort need be expended determining preweighting functions. Instead
these can be obtained for `free' by virtue of histogram reweighting as
we shall now describe.

One starts off by gauging the approximate location of the critical point
from a series of short runs on a small system. This is not time
consuming and can be performed interactively at the computer, provided
one arranges that the data list (see section~\ref{sec:practice}) is
delivered directly to the screen.  One starts by nominating a
temperature $T$ and a large negative value of $\mu$ at which a short
simulation is performed without multicanonical preweighting. The
fluctuating number density will typically settle down quite rapidly, and
its average value can be estimated visually from the data output. One
then repeats this procedure for a succession of progressively larger
$\mu$ values. As $\mu$ is increased, the average particle number will be
observed to increase steadily. However on traversing the
hysteresis-shifted phase boundary, a sudden jump will occur in the
density. If this jump is large, eg. for the LJ fluid, from $\rho=0.05$ to
$\rho=0.6$ (in units of $\sigma^{-3}$), then the temperature is well below
the critical temperature and one should increase $T$ somewhat and begin
again. If the jump is somewhat smaller e.g. from $\rho\simeq 0.2$ to $\simeq
0.4$, then one has obtained an estimate for a near-critical point on the
phase boundary $\mu_\sigma(\beta)$. Of course, if no jump in density is
observed at all then the temperature probably exceeds the critical
temperature and should be reduced.

One next performs a longer run for some larger system of interest at the
estimated near-critical phase boundary point, let us call it
$\mu_\sigma(\beta_0)$. Since the surface tension and the associated
barrier to inter-phase crossings are low near criticality, it should be
possible to accumulate an accurate form for $p(N,E)$ (including
information on states `between the peaks') without resort to
multicanonical preweighting. Having done this, the next step is to
reweight the data accumulated from this run to obtain an estimate of the
form of $p(N)$ at some lower temperature point on the phase boundary \cite{FOOT4}.
This is achieved by first choosing an extrapolation temperature
$\beta_1$ inside the range of reliable reweighting (so that the
reweighted  distribution $p_{ex}(N\;|\;\beta_1,\mu_1)$ appears smooth).
One then tunes $\mu$ within the histogram reweighting scheme until 
$p_{ex}(N\;|\;\beta_1,\mu_1)$ is bimodal with equal area under each
peak. This tuning procedure can be easily automated within the analysis
program to deliver precise values of the phase boundary chemical
potential $\mu_1=\mu_\sigma(\beta_1)$.

The reweighted phase boundary histogram $p_{ex}(N\;|\;\beta_1,\mu_1)$
will, on account of the lower temperature, be more strongly peaked than
that from which it derives. Thus multicanonical preweighting will
probably be necessary for a new simulation at $\beta_1,\mu_1$.
Fortunately however, a suitable preweighting function is already to
hand --it is just the extrapolated function
$p_{ex}(N\;|\;\beta_1,\mu_1)$. Thus all one need do is set
$\eta(N)=p_{ex}(N\;|\;\beta_1,\mu_1)$ and perform a multicanonical
simulation at $\beta_1,\mu_1$ to obtain the actual distribution
$p(N,E\;|\;\beta_1,\mu_1)$. 

One then simply iterates this procedure: histogram reweighting of 
$p(N,E\;|\;\beta_1,\mu_1)$ is used to estimate a phase boundary point
$\mu_2=\mu_\sigma(\beta_2)$ at temperature $\beta_2$, together with the
extrapolated distribution $p_{ex}(N\;|\;\beta_2,\mu_2)$. The latter
serves as a preweighting function for a further simulation at $\mu_2,
\beta_2$, and so on. In this way one steps down the coexistence curve,
obtaining at the same time the locus of the phase boundary
$\mu_\sigma(\beta)$ and the associated set of number density pdf's.
Clearly the maximum feasible step size is set by the range of reliable
histogram extrapolation, which does decreases with increasing system
size. In practice, however, quite large strides can be made even for
large systems. For example, in reference~\cite{WILDING95}, a system of
approximately $600$ LJ particles was studied and $7$ steps were required
to reach a subcritical temperature of $T=0.85T_c$. 

\begin{figure}
\setlength{\epsfxsize}{7.0cm}
\centerline{\mbox{\epsffile{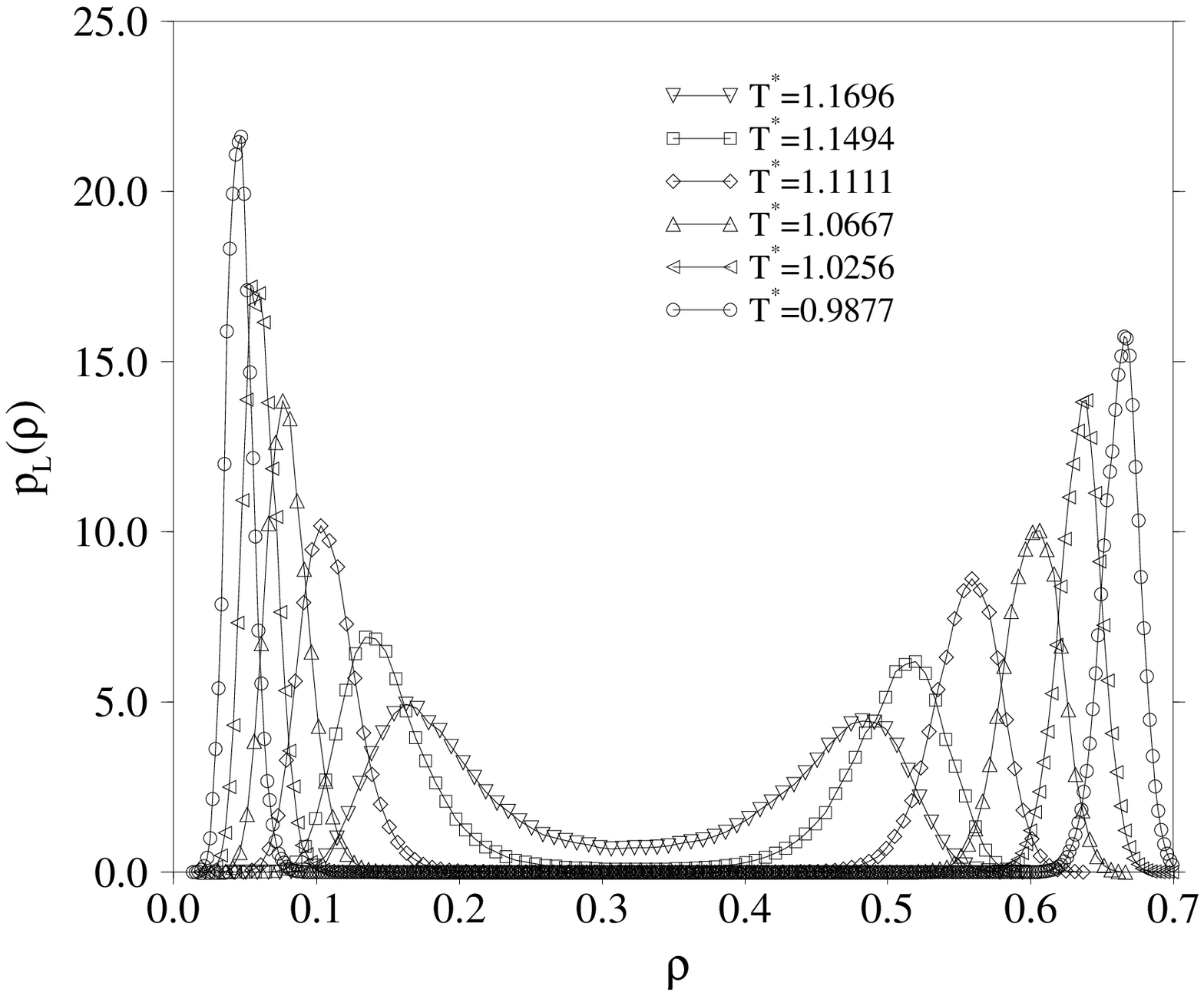}}}
\setlength{\epsfxsize}{7.0cm}
\centerline{\mbox{\epsffile{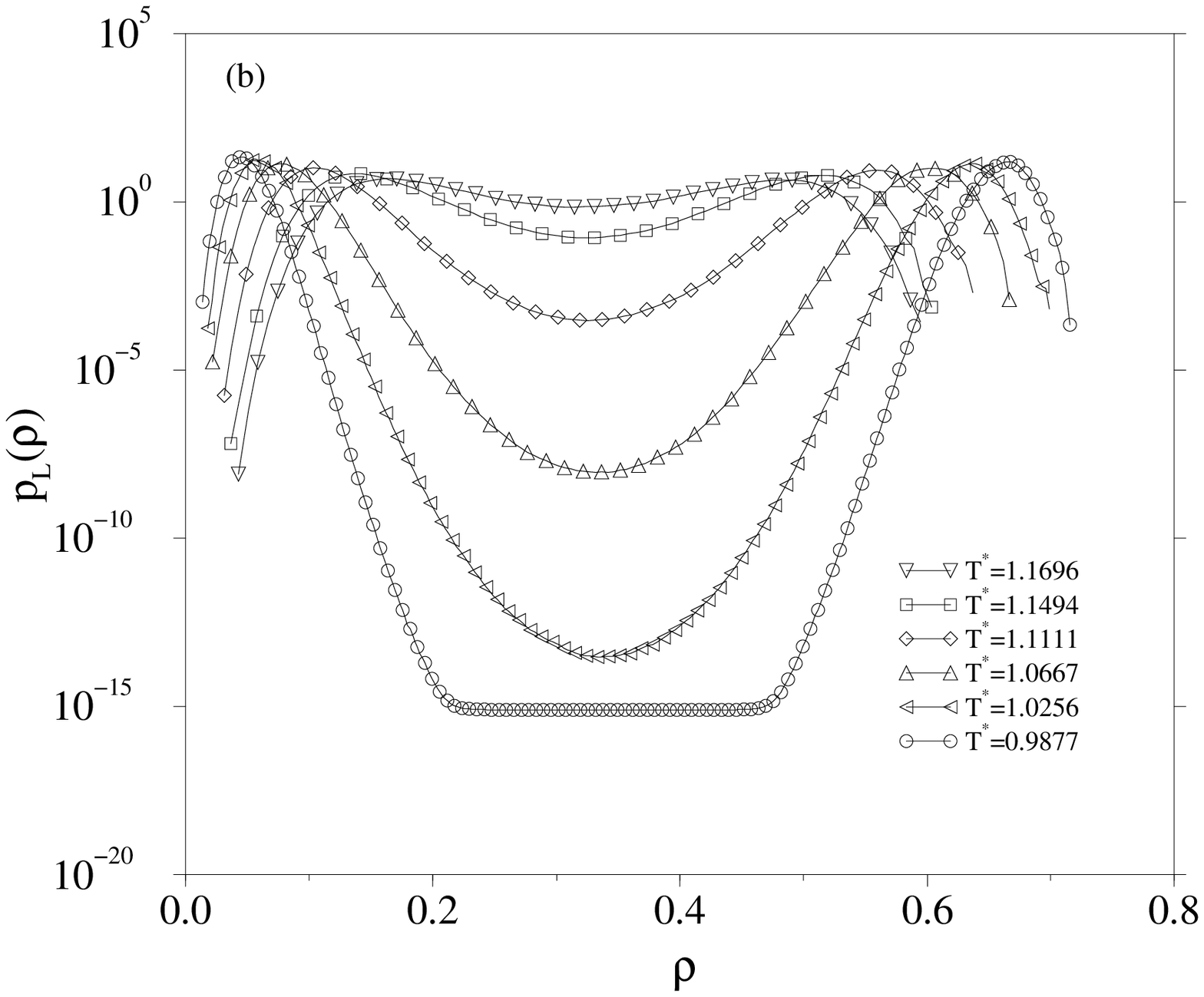}}}

\caption{{\bf (a)} A selection of the measured forms of $p(\rho)$ on the phase
boundary, for temperatures in the range $0.95 \leq T \leq T_c$. {\bf
(b)} The same data expressed on a log scale.}

\label{fig:coexdists}
\end{figure}

\begin{figure}
\setlength{\epsfxsize}{7.0cm}
\centerline{\mbox{\epsffile{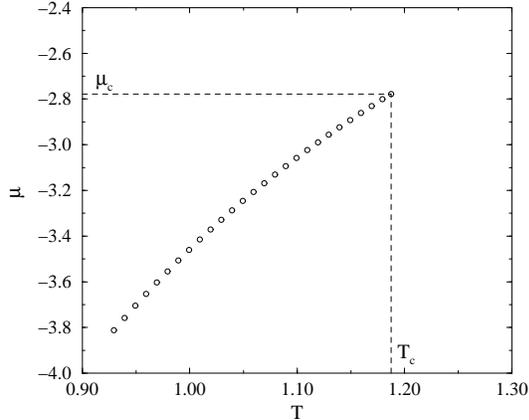}}}
\caption{The line of liquid--vapour phase coexistence in the
space of $\mu$ and $T$, for temperatures in the range $0.95 \leq
T \leq T_c$. Also shown is the location of the critical point\protect\cite{WILDING95}}.
\label{fig:muT}
\end{figure}

The results of implementing this procedure for the LJF \cite{WILDING95}
are depicted in fig.~\ref{fig:coexdists}. The forms for $p(N)$ shown
have equal weight in each peak and hence lie on the phase boundary. The
associated phase diagram $\mu_\sigma(\beta)$ is shown in
figure~\ref{fig:muT}. The values of the coexisting densities can be
simply read off from the peak positions in fig.~\ref{fig:coexdists}(a).
The enormity of the probability barrier that multicanonical preweighting
allows one to negotiate is revealed by plotting these distributions on a
log scale, fig.~\ref{fig:coexdists}(b).

Finally in this section we point out that the measured coexistence form
of the density distribution $p(\rho)$ permits an estimate of the
surface tension $\gamma$. For a cubic system of volume $L^3$,  this is
found simply from the ratio of values of $p(\rho)$ at the peak and at
the trough between the peaks \cite{BINDER82}:

\[
\gamma=\frac{1}{2\beta L^2}\ln\left(\frac{p_{\rm max}}{p_{\rm min}}\right)
\]
Reference~\cite{POTOFF00} describes a recent application of this relation in a
study of the surface tension of the Lennard-Jones fluid.

\section{Discussion and conclusions} 

\label{sec:concs}

To summarise, we have described a scheme whereby information on the
locus of a fluid phase boundary is obtainable via the dual mean of
multicanonical preweighting and histogram reweighting, married to a
standard grand canonical simulation algorithm. The method begins tracing
the phase boundary from near the critical point where the free energy
barrier to inter-phase transitions is small. Histogram reweighting of
the data thus obtained is used to provide an estimate of the location of
the phase boundary at some lower temperature, together with a suitable
form of the requisite preweighting function. A new multicanonical
simulation is performed at this lower temperature phase boundary point
and the process is repeated. In this way it is possible to stride down
the phase boundary, obtaining pdfs of observables such as the number
density from which, in turn, the coexistence properties can be inferred.
Such use of multicanonical preweighting completely eliminates 
hysteresis at first order phase transition.

In cases where one doesn't wish to start tracing a coexistence curve
from near the critical point, it is necessary to bootstrap the procedure
described above by obtaining an initial phase boundary preweighting
function. A variety of techniques exist for achieving this, ranging from
simple extrapolation of the weight function into the unsampled region, to
more sophisticated analyses of MC transition probabilities. Most of the
techniques in common use can be straightforwardly automated. It is beyond
the scope of this article to describe these methods in detail, and for
further details we refer the interested reader to the
literature~\cite{SMITH,GUBERNATIS}.

Although we have illustrated our approach solely in the context of a
simple fluid model, it should be noted that it is equally applicable
to complex fluids such as molecules or polymers. In these systems,
however, the GCE insertion probability is often small, so it is necessary
to supplement the standard algorithm with a more intelligent insertion
scheme --one which performs a biased choice of a molecular orientation
favourable for the insertion. Methods such as Configurational Bias Monte
Carlo \cite{FS97} and Recoil Growth \cite{CONSTA99} allow one to do
this. Apart from this added complication, multicanonical preweighting
and histogram reweighting are implemented exactly as for a simple fluid.

The scheme we have described is generalisable to other simulation
ensembles such as the constant-NpT ensemble in which density
fluctuations occur by means of volume changes at constant particle
number, pressure and temperature. Use of this ensemble can be more
efficient than the GCE when dealing with very dense fluids where the
success rate of particle insertions and deletions is small. The
natural variable in which to preweight a constant-NpT ensemble
simulation is the fluctuating volume, but otherwise the formalism is
very similar to that described above. For other types of phase
transitions, such as the liquid-liquid transitions occurring in binary
fluid mixtures, a suitable variable in which to preweight is usually the
order parameter for the transition eg. the concentration of one species.

It is instructive to compare the scheme described in this article with
other commonly used methods for tracing phase boundaries. One such
method is the Gibbs-Duhem integration method wherein pairs of single
phase simulations (one for each phase) are performed at various state
points along the phase boundary.  The single phase results are connected
thermodynamically, not by negotiating the free energy barrier at each
state point, but via a phase space path that runs back along each side
of the phase boundary to some independently-known reference point on the
phase boundary. Successive simulation state points along the phase
boundary are found from an integration scheme which turns out
\cite{ESCOBEDO} to be a low order approximation to the histogram
reweighting method. The Gibbs-Duhem method is versatile provided one has
prior knowledge of a phase boundary reference point and is, for the
purposes of obtaining a rough estimate of the phase boundary locus,
doubtless faster than the method described in this article. However, its
accuracy rests heavily upon the precision with which the boundary
reference point is known. Moreover, as one departs from this point,
integration errors can grow and provided one remains within the (rather
wide) region of metastability bordering either side of the phase
boundary line, there is no feedback to indicate that one has in fact
strayed away from it.  By contrast, the multicanonical method is
self-correcting because it reconnects the two phases at each successive
simulation state point on the phase boundary. 

Another widely used technique is the Gibbs Ensemble MC Method 
\cite{PANAGIO87}. Two separate simulation boxes (one for each pure
phase) are connected thermodynamically by the dual means of particle
transfers between the boxes and fluctuations in the box volumes
implemented such that the overall volume of the two boxes remains
constant. Like Gibbs-Duhem integration, GEMC is very useful for
obtaining rough estimates of phase boundaries.  The method is free of
integration errors since the phases remain directly linked at each phase
boundary state point. However, measurements of the pressure and chemical
potential must be obtained by a sampling scheme since neither is imposed
in the simulation. Moreover, direct comparisons have shown GEMC to be
much less efficient than the methods described in this article because
of the computational complexity associated with implementing volume
fluctuations \cite{POTOFF99,PANAGIO00}.  Worries have also recently been
voiced that the GEMC method suffers more severely from finite size
effects than does the GCE \cite{VALLEAU98}.

Finally, it should be pointed out that despite their utility in dealing
with phase equilibria involving fluids, the specific multicanonical
techniques we have described do not permit one to tackle phase
transitions involving solids. The problem here is that when attempting
to traverse mixed-phase states involving crystalline order, the
simulation invariably gets caught in `non-ergodic traps' identifiable
with defective crystalline configurations. Recently, however, a new
technique has been developed that circumvents this problem by linking
the two coexisting phases {\em without} traversing mixed-phase states.
Essentially the method can be thought of as leaping directly from the
configuration space of one pure phase to that of the other. Again use of
multicanonical sampling is necessary, but this time its role is to
encourage the simulation to visit a subset of configurations (in each
pure phase) from which a leap to the other phase will be accepted. This 
new method has recently shown its worth in studies of solid-phase free
energy differences and hard sphere freezing \cite{BRUCE97,WILDING00}.

\section{Suggestions for further study}

The following set of problems will help to reader to become more
familiar with the techniques described in this article.

\begin{enumerate}

\item Using Box 1. as a guide, write a simple Grand Canonical ensemble
MC program to simulate the Lennard-Jones fluid with potential cut-off at
$r_c=2.5\sigma$. Do not correct for the potential truncation.  Make sure
the program prints out the particle number and the energy in list form
(cf. sec.~\ref{sec:practice}). Further programming details can be found
in ref.\cite{FRENKEL}.

\item Run the program at the near critical phase boundary parameters
\cite{WILDING95} $\beta\epsilon=1.1876$, $\beta\mu=-2.778$, saving the
output data list to a file.

\item Write a post-processing program to construct the number density
histogram $p(N)$ from the raw data list.

\item Modify your GCE acceptance probabilities to cater for
multicanonical preweighting (cf. secs.~\ref{sec:multi} and
~\ref{sec:practice}). Use your measured $p(N)$ as the preweighting
function for a multicanonical simulation at the same $\beta,\mu$ used in
(2) above. Hint: At the extrema of small (large) $N$, there will be
histogram bins in $p(N)$ having zero entries. Before using $p(N)$ as
your preweighting function, set these entries to be a constant equal to
the smallest non zero entry.  This avoids possible division by zero in
the acceptance probabilities.

\item Extend your post-processing program to unfold the effects of the
preweighting (as described in Secs.~\ref{sec:multi} and
\ref{sec:practice}) in order to find $p(N)$. Check that the form of $p(N)$ thus
obtained agrees with that found without multicanonical preweighting.
Compare the correlation time for the sampling
processes with and without multicanonical preweighting.

\item Further extend your post-processing program to implement histogram
reweighting (as described in Sec.~\ref{sec:histo} and
\ref{sec:practice}). Extrapolate the data obtained at the critical
temperature $\beta\epsilon=1.1876$ to find the location of the phase
boundary and the form of $p(N)$ at $\beta\epsilon=1.17$. Use this
extrapolation as a preweighting function in a new multicanonical
simulation at the new phase boundary state point. Compare your results
with those of ref.~\cite{WILDING95}

\end{enumerate}

%\end{multicols}

\begin{thebibliography}{99}
\bibitem{FOOT0} A comparison of the model's phase behaviour with the
results of experiment can be useful in helping to refine the model
parameters.
\bibitem{FOOT2}  The critical point is characterised by strong
fluctuations in the density which present distinctive challenges for
simulation and theory alike. For an introduction to some of the main
issues, see chapter 5 of ref.~\protect\cite{PLISCHKE} and also
ref.~\cite{WILDING95}.
\bibitem{FOOT3} In fact it is possible to preweight with respect to any
system observable, although for the problem of fluid phase transitions
the number density is usually the most computationally appropriate
quantity.
\bibitem{FOOT4} One can also extrapolate to slightly higher temepratures
in order to pinpoint the critical point. Special `finite-size scaling'
techniques are required here, as described in detail in
reference~\protect\cite{WILDING95}.
\end{thebibliography}
\end{document}